# Intersecting topological nodal ring and nodal wall states in superhard superconductor FeB$_4$

Feng Zhou[a], Ying Liu[b]*, Jianhua Wang[a], Tie Yang[a], Hong Chen[a], Xiaotian Wang[a]*, and Zhenxiang Cheng[c]*

[a]*School of Physical Science and Technology, Southwest University, Chongqing 400715, China;*

[b]*School of Materials Science and Engineering, Hebei University of Technology, Tianjin 300130, China*

[c]*Institute for Superconducting and Electronic Materials (ISEM), University of Wollongong, Wollongong 2500, Australia*

*Corresponding author.*

*email addresses: ying_liu@hebut.edu.cn; xiaotianwang@swu.edu.cn; cheng@uow.edu.au*

**Abstract**

Novel materials with both topological nontrivial states and superconductivity have attracted considerable attention in recent years. Single-crystal FeB$_4$ was recently synthesized and demonstrated to exhibit superconductivity at temperatures lower than 2.9 K, and its nanoindentation hardness was measured to be 65 GPa. In this study, based on first-principles calculation and the low-energy $k \cdot p$ effective Hamiltonian, we found that this *Pnnm*-type superhard FeB$_4$ superconductor hosts topological behaviors with intersecting nodal rings (INRs) in the $k_x = 0$ and $k_z = 0$ planes and nodal wall states in the $k_y = \pi$ and $k_z = \pi$ planes. The observed surface drum-head-like (D-H-L) states on the [100] and [001] surfaces confirmed the presence of INR states in this system. According to our investigation results, FeB$_4$, with its superconductivity, superior mechanical behaviors, one-dimensional (1D) and two-dimensional topological elements, and D-H-L surface states, is an existing single-phase target material that can be used to realize the topological superconducting state in the near future.

I. **Introduction**

Topological superconductors (TSCs)[1-5] can be used to achieve Majorana bound states and other exotic physical properties, for example, emergent supersymmetry. Therefore, in recent years, the quest to design and search for new TSCs has emerged as one of the most important directions in condensed matter physics and quantum chemistry. In general, two approaches are followed to design topological superconductors. The first approach[6] involves constructing heterostructures consisting of topological materials (TMs) and superconducting materials (SCMs), followed by use of the proximity effect to achieve topological superconductivity. In this approach, one should consider the lattice mismatch and interface reactions between the two materials, which usually hinder successful construction of the heterostructure. The second approach[7] involves achieving topological superconductivity in a single-phase material instead of a heterostructure. In this approach, one can select an SCM or a TM as the starting material platform and subsequently induce the topological property or superconductivity in the said material.

To date, a few topological insulators and topological semimetals that feature nodal points and superconductivity have been reported. For example, $Bi_2Se_3$[8] is a well-known topological insulator, and Hor et al.[9] reported that Cu intercalation in the gaps between two $Bi_2Se_3$ layers can induce superconductivity at 3.8 K in $Cu_xBi_2Se_3$, where $x$ = 0.12–0.15. Similarly, in 2015, Liu et al.[10] reported the existence of superconducting and surface topological states in $Sr_{0.065}Bi_2Se_3$. In $Cd_3As_2$[11] and tungsten ditelluride[12] materials with a zero-dimensional (0-D) topological element (TE), superconductivity can be induced by applying high pressure. Furthermore, $MoTe_2$[13], a sister compound of tungsten ditelluride, was proved to exhibit the Weyl semimetallic property and to intrinsically host superconductivity with a transition temperature ($T_c$) of 0.10 K. In 2019, based on first-principles calculations, Song et al.[14] predicted that highly stable two-dimensional (2D) rect-$AlB_6$ material features triple Dirac cones, a remarkable motif, and possibly an intrinsic superconducting character.

Recently, researchers have also been keen to develop materials that feature both topological nodal lines (NLs) and superconducting states. Thus far, a handful of known TMs with a one-dimensional (1D) TE accompanied by superconductivity have been predicted, including NaAlSi[15], [Ti/Mg]Bi$_2$[16,17], [Sn/Pb]TaSe$_2$[18,19], SrAs$_3$[20,21], and In$_x$TaS$_2$[22]. To the best of our knowledge, TMs with a 2D TE, namely, nodal wall (also called nodal surface)[23-25] states and possible superconductivity, have not been analyzed by researchers. Therefore, the search for highly stable, purpose-built TMs with simultaneous 1D/2D TE and superconducting states is highly significant.

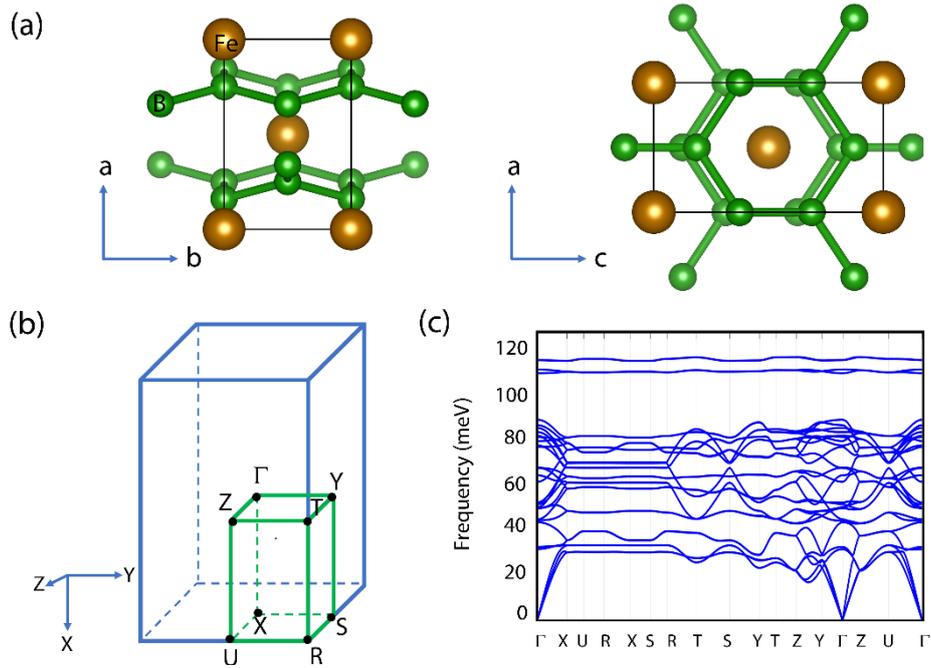

Fig. 1 (a) Crystal structures of *Pnnm*-type FeB$_4$ system from different views; the green and brown spheres represent B and Fe atoms, respectively; (b) bulk Brillouin zone (BZ); and (c) calculated phonon dispersion of the FeB$_4$ system along the Γ-X-U-R-X-S-R-T-S-Y-T-Z-Y-Γ-Z-U-Γ paths.

In the present work, we study the topological property of a superhard iron tetraboride superconductor, namely, FeB$_4$, with the *Pnnm*-type structure[26]. Gou et al[26] synthesized single crystals of *Pnnm*-type FeB$_4$ at pressures higher than 8 GPa and high temperatures. Interestingly, this system[26] has been proved to exhibit bulk superconductivity at temperatures lower than 2.9 K based on magnetic susceptibility measurements. Moreover, FeB$_4$ has high levels of hardness[26], that is, no phase transitions occur at ambient temperature in a diamond anvil cell under pressures of up to ca. 40 GPa. Hence, this superhard FeB$_4$ superconductor can be viewed as an excellent target material to study the perfect mechanical and superconductivity properties[26]. Based on first-principles calculations, we reveal that this prepared compound is also a topological material with 1D nodal rings (NRs) and 2D nodal walls. Such a superhard material could be a good target material to further investigate the entanglement between the topological and the superconducting states. Importantly, we, for the first time, report superconductor co-hosts 1D and 2-D TEs, thus demonstrating that FeB$_4$ is a suitable material for achieving topological superconductivity.

## II. Materials and Computational Details

The band structures and topological signatures were obtained using the VASP code[27]. The exchange-correlation potential was treated using the generalized gradient approximation (GGA)[28] with the Perdew–Burke–Ernzerhof (PBE)[29] functional. The cutoff energy was set to 600 eV. A Monkhorst–Pack special 11 × 11 × 7 $k$-point mesh was used in the BZ integration. The surface states of FeB$_4$, including the [100] and [001] surface states, were investigated using the WannierTools software application[30]. The phonon dispersion of FeB$_4$ was calculated using the Nanodcal code[31] and the force-constants method.

The optimized crystal structures of *Pnnm* FeB$_4$ are shown in Fig. 1(a), where the Fe atoms and B atoms occupy the 2a (0.5, 0.5, 0.5) and 8h (0.5, 0.75, 0.19) Wyckoff positions, respectively. The optimized lattice constants are $a$ = 3.00 Å, $b$ = 4.503 Å, and $c$ =5.278 Å. The obtained values are close to the experimental data of FeB$_4$ [26] ($a$ = 2.99 Å, $b$ = 4.578 Å, and $c$ =5.298 Å). The phonon dispersion of the FeB$_4$ system along the Γ-X-U-R-X-S-R-T-S-Y-T-

Z-Y-Γ-Z-U-Γ paths (see Fig. 1(b)) is shown in Fig. 1(c). The lack of imaginary frequencies in the phonon dispersion indicates the dynamic stability of this system.

## III. Intersecting Nodal Ring and Nodal Wall States

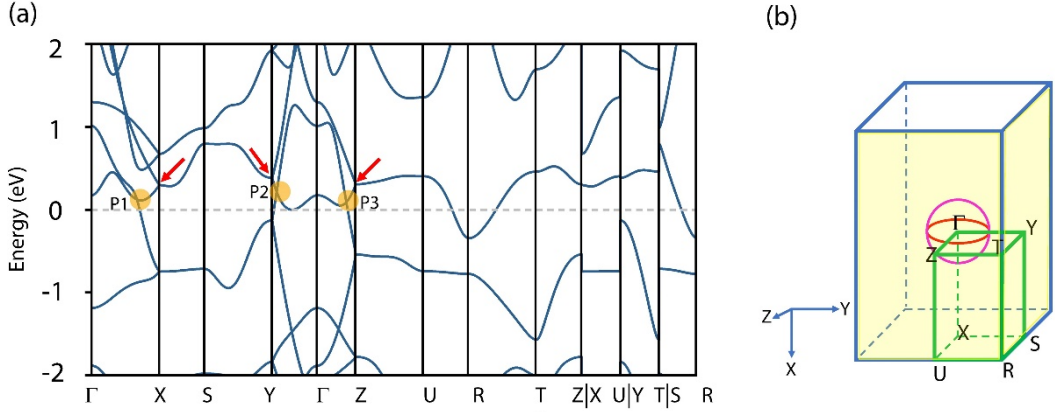

Fig. 2 (a) Band structure of FeB$_4$ along the Γ-X-S-Y-Γ-Z-U-R-T-Z|X-U|Y-T|S-R paths. The BCPs around the Fermi level are labeled P1, P2, and P3; (b) schematic diagram of INRs, in the $k_x = 0$ and $k_z = 0$ plane, and nodal walls in the $k_y = \pi$ and $k_z = \pi$ planes.

Fig. 2(a) shows the band structure of FeB$_4$ without the spin-orbit coupling. FeB$_4$ is a metallic system with three clear band-crossing points (BCPs)[32,33], one each along the Γ-X, Y-Γ, and Γ-Z paths, named P1, P2, and P3, respectively. Given that Fe is a transition metal, the DFT+$U$ calculation[34,35], where $U$ is the Hubbard correction, was performed to examine its electronic structures (see Fig. S1). In Fig. S1, one can see three clear BCPs around the Fermi level (E$_F$) that persist when $U$ = 2-5 eV. Because FeB$_4$ has $P$ and $\mathcal{T}$ symmetries, points P1, P2, and P3 should not be isolated, as argued by Weng et al[36-39]. To confirm the nodal states of the FeB$_4$ system, we performed a detailed scan of the band structures along the selected $k$-paths in the $k_{x,y,z} = 0$ planes (see Fig. 3(a), Fig. 4(a), and Fig. 5(a)).

The band structures along the abovementioned $k$-paths are shown in Fig. 3(b), Fig. 4(b), and Fig. 5(b). The BCPs persist along all of the Γ-a1/a2/a3/a4, Γ-b1/b2/b3/b4, Γ-c1/c2/c3/c4, and Γ-d1/d2/d3/d4 paths, indicating that the closed NR states exist in the $k_x = 0$ and $k_z = 0$ planes. In addition, we plot the 3D Γ-centered band structures and the shapes of the closed NRs of both planes in Fig. 3(c) and Fig. 4(c), and Fig. 3(d) and Fig. 4(d), respectively. However, in the band structures along the *k*-paths, namely, Γ- e1/e2/e3/e4 and Γ- f1/f2/f3/f4, of the $k_y = 0$ plane, no BCPs occur (see Fig. 4b). Moreover, the two NRs in the $k_x = 0$ and $k_z = 0$ planes share the same center point Γ, reflecting that FeB$_4$ is a TM with intersecting nodal rings (INRs)[40-43] in the momentum space (see Fig. 2(b)). The INRs are extremely close to the E$_F$, and the energies of these two closed NRs vary within small ranges (see Fig. 3(b) and Fig. 4(b)).

To further discuss the INR states of this system, we attempt to understand them based on the effective Hamiltonian. The little group at point Γ belongs to the D$_{2h}$ generated by M$_x$, M$_y$, and M$_z$, and it exhibits C$_{2x}$, C$_{2z}$, and inversion symmetry. The nodal ring on the k$_x$ = 0 plane is protected by M$_x$, and the two crossing bands are characterized by the eigenvalues, namely, $g_z = 1$, and $g_z = -1$, respectively. Thereby, the mirror symmetry may take the matrix form as

$$M_x = \sigma_z \tag{1}$$

and satisfy

$$M_x H(k) M_x^{-1} = H(-k_x, k_y, k_z). \tag{2}$$

Together with the remaining symmetries, we have

$$C_{2z} H(k) C_{2z}^{-1} = H(-k_x, -k_y, k_z), \tag{3}$$

$$P H(k) P = H(-k_x, -k_y, -k_z). \tag{4}$$

Consequently, the effective Hamiltonian can be expressed as

$$H(k) = w(k)\sigma_0 + [M + Ak_x^2 + Bk_y^2 + Ck_z^2]\sigma_z + \gamma k_x k_z \sigma_x. \tag{5}$$

In the same way, the nodal ring on plane $k_z = 0$ takes the same form. Equation (5) indicates that on plane $k_x = 0$ or $k_z = 0$, the term $\gamma k_x k_z$ vanishes. The NRs on these two planes are determined by the diagonal term, which is proportional to $\sigma_z$. For the nodal ring on $k_x = 0$, we have $h_{k_x=0}(0, k_y, k_z) = M + Bk_y^2 + Ck_z^2 = 0$, where *MB* < *0* and *MC* < *0*, which leads to 1D band crossing between the two bands on this plane. Referring to the nodal ring on plane $k_z = 0$, the band crossing is determined by $h_{k_z=0}(k_x, k_y, 0) = M + Ak_x^2 + Bk_y^2 = 0$, also with *MB* < *0* and *MA* < *0*, thus establishing a ring. Therefore, it is proved that there are two NRs, as shown in Fig. 2(b).

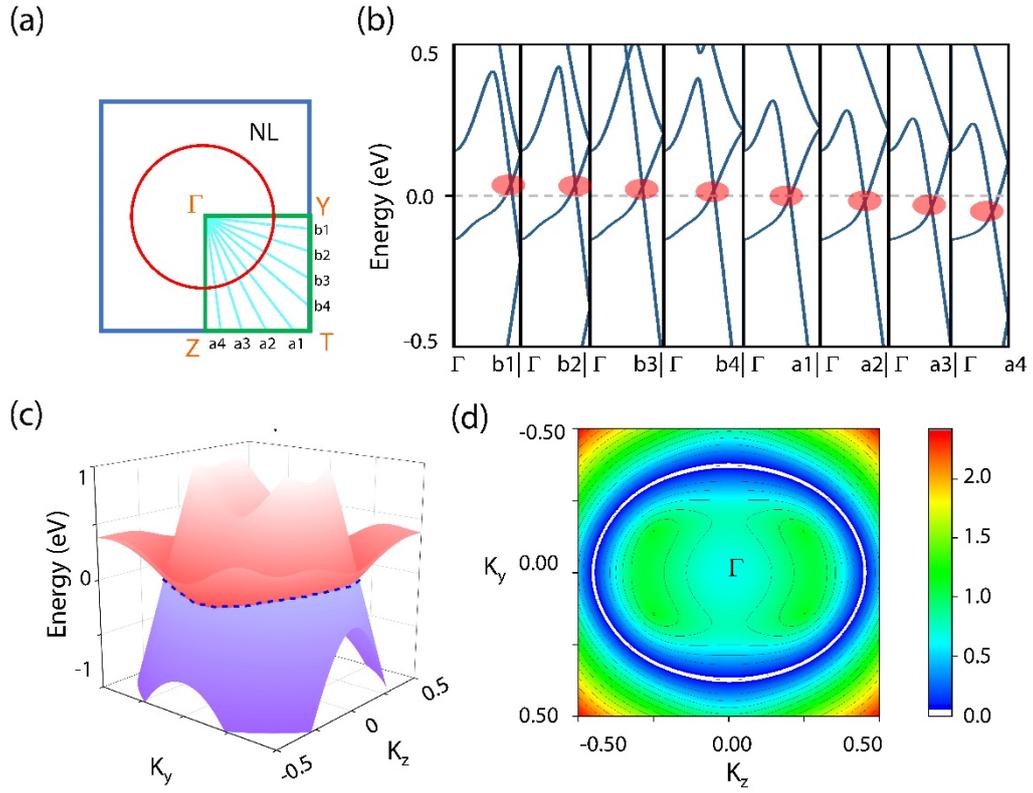

Fig. 3 (a) Selected $k$-paths on the $k_x = 0$ plane. The points a1-a4 (b1-b4) are equally spaced between T and Z (Y and T); (b) the band structures of the Γ-a1/a2/a3/a4 and Γ- b1/b2/b3/b4 paths in the $k_x = 0$ plane; (c) 3D band dispersion of the $k_x = 0$ plane around point Γ. (d) Shapes of the Γ-centered NR of the $k_x = 0$ plane. The dotted and white solid lines in (c) and (d) represent the NR states in the $k_x = 0$ plane.

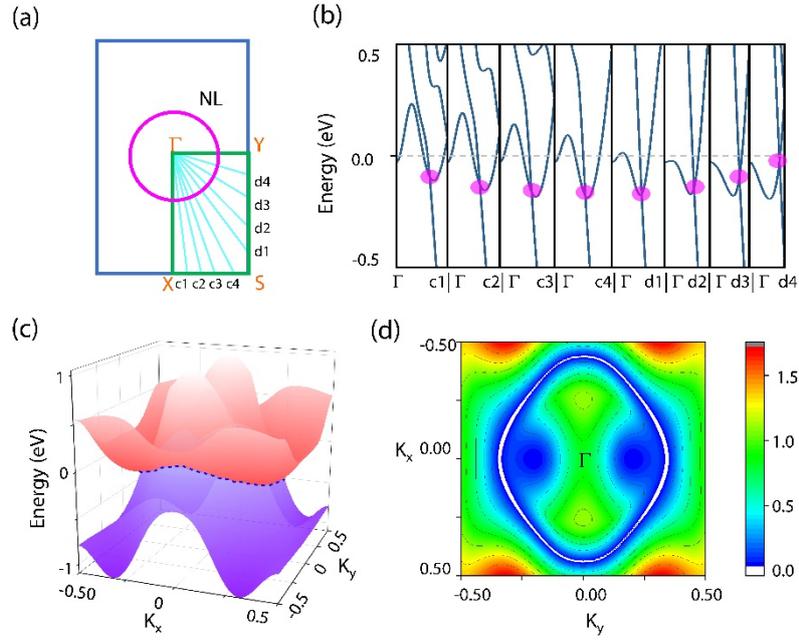

Fig. 4 (a) Selected *k*-paths of the $k_z = 0$ plane, where the points c1–c4 (d1–d4) are equally spaced between X and S (S and Y); (b) band structures of the Γ-c1/c2/c3/c4 and Γ- d1/d2/d3/d4 paths in the $k_z = 0$ plane; (c) 3D band dispersion of the $k_z = 0$ plane around point Γ. (d) Shapes of the Γ-centered NR of the $k_z = 0$ plane. The dotted and white solid lines in (c) and (d) represent the NR states in the $k_z = 0$ plane.

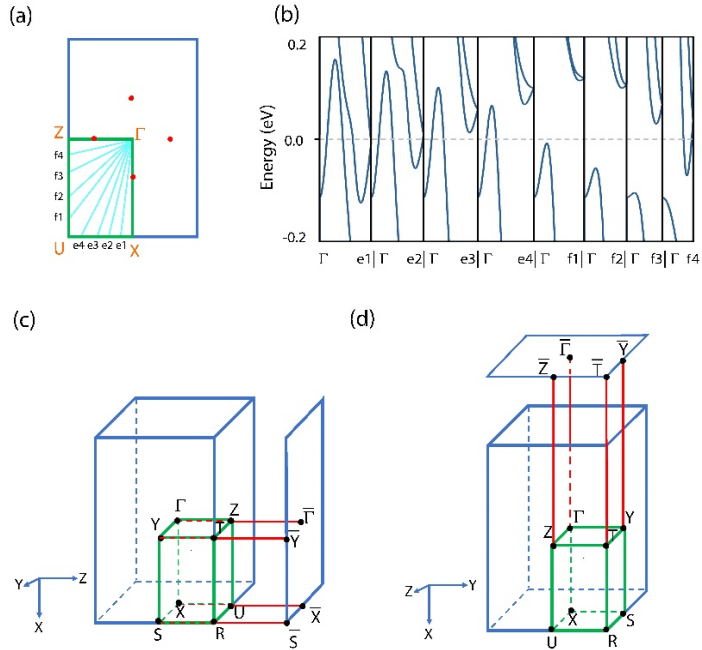

Fig. 5 (a) Selected *k*-paths in the $k_y = 0$ plane, where the points e1–e4 (f1–f4) are equally spaced between X and U (U and Z); (b) band structures of the Γ-e1/e2/e3/e4 and Γ- f1/f2/f3/f4 paths in the $k_y = 0$ plane; (c) and (d) bulk BZ and surface BZ of the [001] surface and [100] surface, respectively.

In addition to the BCPs P1, P2, and P3, in Fig. 2(a), two bands linearly cross at the X, Y, and Z points (see the red arrows). However, along the paths, X-S-Y and Z-U-R-T-Z|X-U|Y-T|S-R, the two bands degenerate, forming a nodal wall state, also called a nodal surface state[23-25,44,45]. We believe that the nodal walls should be appeared in the $k_y = \pi$ and $k_z = \pi$ planes. To further confirm the nodal walls of these two planes, we performed a detailed electronic-structures calculation based on the selected $k$-paths in the $k_y = \pi$ and $k_z = \pi$ planes. The selected $k$-paths are shown in Fig. 6(a), and the calculated band structures along the f1-h1/f2-h2/f3-h3/f4-h4/f5-h5/f6-h6/f7-h7/f8-h8 paths are shown in Fig. 6(b). From Fig. 6(b), one finds that all of the band structures are degenerated in both planes, indicating the presence of nodal wall states. Moreover, the nodal wall states are closed along the $k_y$ and $k_z$ directions (see Fig. 2(b)).

Notably, the closed nodal wall/surface states along the $k_y$ and $k_z$ directions are essential, and their formation can be analyzed in terms of symmetry. When SOC is not included, such surfaces are enforced by the combination of a twofold screw symmetry and time reversal symmetry ($\mathcal{T}$). Indeed, there exist two screw rotations perpendicular to each other, namely, $S_{2y}: \left(-x + \frac{1}{2}, y + \frac{1}{2}, -z + \frac{1}{2}\right), S_{2z}: \left(-x + \frac{1}{2}, -y + \frac{1}{2}, z + \frac{1}{2}\right)$. Accordingly, we have $(S_{2y})^2 = T_{010} = e^{ik_y}$ and $(S_{2z})^2 = T_{001} = e^{ik_z}$, where $T_{010}$ and $T_{001}$ are the translations along the y and z directions, respectively. Because $\mathcal{T}$ is an antiunitary operator and inverses the momentum $\mathbf{k}$, the combined operator $\mathcal{T}S_{2i}$, ($i = y, z$) is antiunitary and only inverse the momentum along the $i$-direction. Furthermore, $\mathcal{T}S_{2i}$ satisfies

$$(\mathcal{T}S_{2i})^2 = e^{-ik_i}, \tag{6}$$

such that the nodal surface can be understood from the Kramer degeneracy. On the corresponding planes $k_y = \pi, k_z = \pi$, we have

$$(\mathcal{T}S_{2y})^2 = -1, \tag{7}$$

$$(\mathcal{T}S_{2z})^2 = -1 \tag{8}$$

on the entire planes. Thus, a double degenerate nodal wall/surface occurs (see Fig. 2).

Although the nodal points and NLs with 0D and 1D band degeneracies have been confirmed in photonics and acoustics[46,47], the nodal wall, that is, the nodal surface state has not been observed in any classical-wave system before 2019. Yang et al.[48], for the first time, confirmed the nodal wall states in a 3D chiral acoustic crystal. Hence, it is hoped that the nodal wall state of FeB$_4$ can be experimentally confirmed soon.

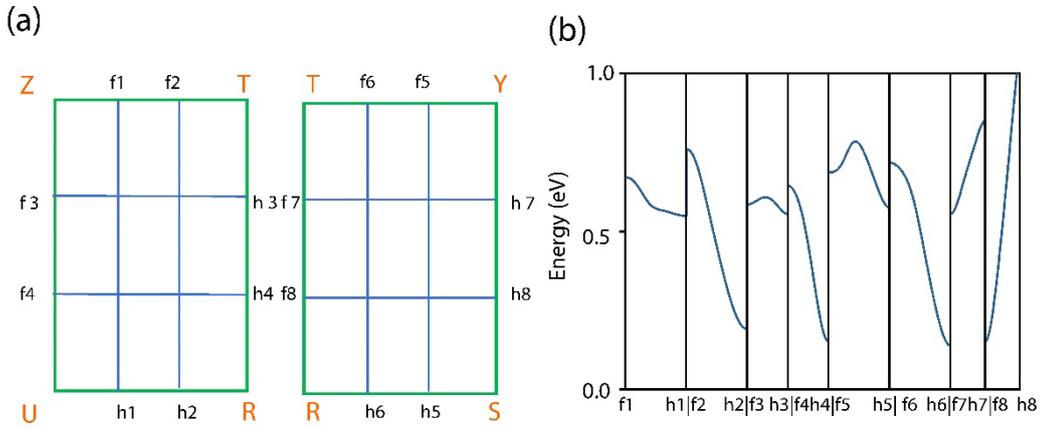

Fig. 6 (a) Selected $k$-paths of the $k_y = \pi$ and $k_z = \pi$ planes; (b) band structures of the f1-h1|f2-h2|f3-h3|f4-h4|f5-h5|f6-h6|f7-h7|f8-h8 paths.

**IV. Surface States and Effect of SOC**

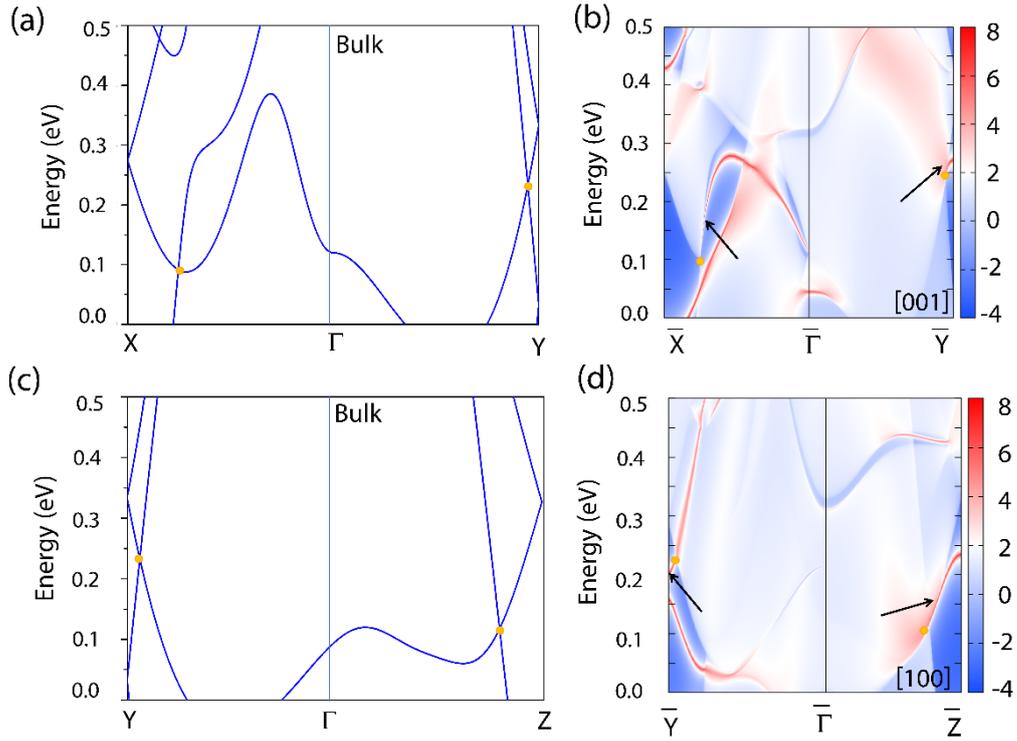

Fig. 7 (a) and (c) Band structures of bulk FeB4 along the X- Γ-Y and Y-Γ-Z paths, respectively. (b) and (b) [001] and [100] surface states of FeB4 along the surface BZ $\bar{X} - \bar{\Gamma} - \bar{Y}$ and $\bar{Y} - \bar{\Gamma} - \bar{Z}$, respectively. The positions of the BCPs are marked by yellow dots, and the drum-head-like surface states arising from the BCPs are marked by black arrows.

For nodal line (NL) materials, we can observe the drum-head-like (D-H-L) surface states arising from the projected bulk NLs[49-53]. In this section, we calculate the projected spectrum of the FeB4 [001] and [100] surface states along the $\bar{X} - \bar{\Gamma} - \bar{Y}$ and $\bar{Y} - \bar{\Gamma} - \bar{Z}$ surface BZ to confirm the appearance of the D-H-L surface states (see Fig. 7(b) and Fig. 7(d)). For comparison, the band structures with marked BCPs along the X- Γ-Y and Y-Γ-Z 3D bulk paths are illustrated in Fig. 7 (a) and Fig. 7(c). Fig. 7(c) and Fig. 7(d) indicate the clear presence of D-H-L surface states on the [001] and [100] surfaces arising from the four BCPs in FeB4. Hopefully, such clear surface states can be observed by using surface-sensitive detection methods.

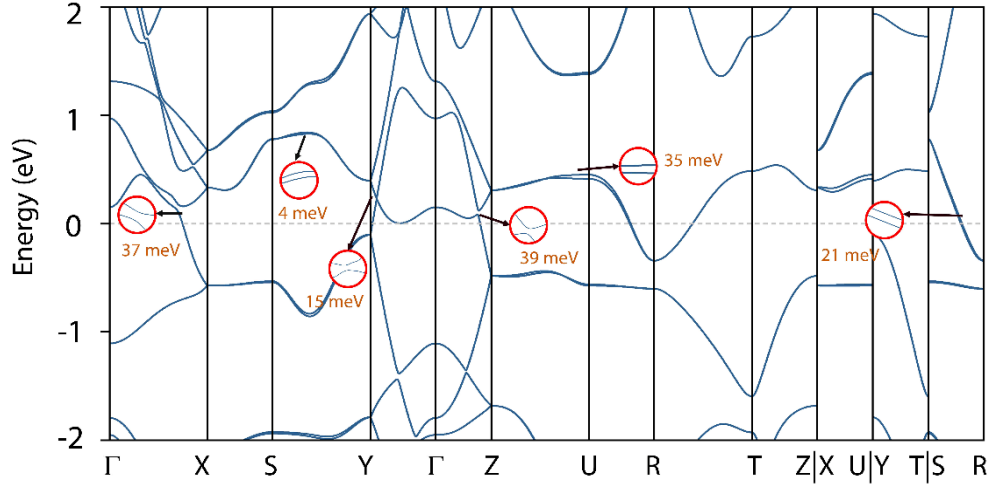

Fig. 8 The calculated band structure of FeB$_4$. SOC is considered.

Finally, we need to the study the effects of SOC on the electronic structures and BCPs of FeB$_4$. The calculated electronic structures of band structures of FeB$_4$ under the SOC effect are shown in Fig. 8. The opened gaps for the NR and nodal wall states are smaller than 39 meV. The values of the SOC-induced gaps in FeB$_4$ are comparable to those in the experimentally confirmed NL material Mg$_3$Bi$_2$ (> 36 meV)[54,55], and they are smaller than those of a few well-known typological materials (with SOC-induced gaps larger than 40 meV), such as CaTe (50–52 meV)[56], Zn$_3$Bi$_2$ (~65.9 meV)[57], Hg$_3$Bi$_2$ (~100.2 meV)[57], Cu$_3$PdN (> 60 meV)[58], CaAs$_3$ (39.92–54.47 meV)[38], and CaAgBi (> 80 meV)[59].

## IV. Conclusions

In this work, we investigated the topological elements in superhard FeB$_4$ superconductor. This system exhibits INR states (centered at point Γ) in the $k_x = 0$ and $k_z = 0$ planes and closed nodal walls (along the $k_y$ and $k_z$ directions) in the $k_y = \pi$ and $k_z = \pi$ planes. Based on the obtained band structures under the SOC effect, we can conclude that the SOC-induced gaps in these topological signatures of FeB$_4$ are comparable to those of an experimentally confirmed NL material Mg$_3$Bi$_2$. Clear D-H-L surface states were found on the [100] and [001] surfaces, reflecting the

1D NR states of the FeB$_4$ bulk material. Our study revealed that FeB$_4$ with the *Pnnm* structure could serve as a novel single-phase topological superconductor material.

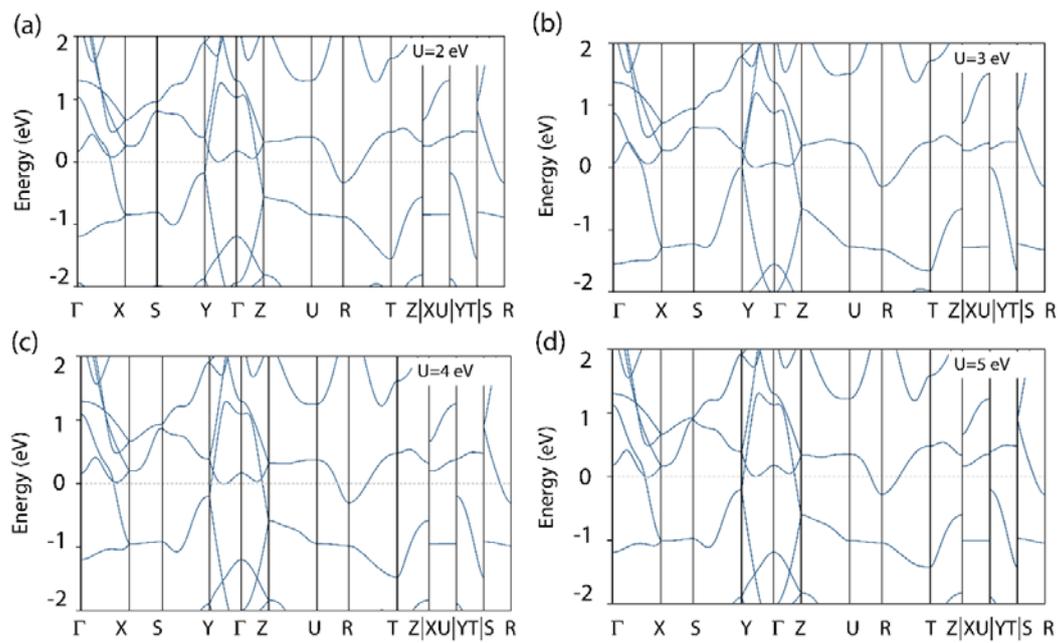

Fig. S1 calculated band structure of FeB$_4$ with GGA+U methods.